# Physical properties of amorphous molybdenum silicide films for single-photon detectors

**Xiaofu Zhang[1,2,3]*, Ilya Charaev[4]†, Huanlong Liu[1], Tony X. Zhou[4], Dong Zhu[1], Karl K. Berggren[4], Andreas Schilling[1]**

[1]Department of Physics, University of Zurich, CH-8057 Zürich, Switzerland
[2]State Key Laboratory of Functional Materials for Informatics, Shanghai Institute of Microsystem and Information Technology, Chinese Academy of Sciences (CAS), Shanghai 200050, China
[3]CAS Center for Excellence in Superconducting Electronics, Shanghai 200050, China
[4]Massachusetts Institute of Technology, 50 Vassar Street, Cambridge, MA 02139, USA

E-mail: *zhang@physik.uzh.ch;†charaev@mit.edu

### Abstract

We systematically investigated the physical properties of amorphous $Mo_xSi_{1-x}$ films deposited by magnetron co-sputtering. The critical temperature $T_c$ of $Mo_xSi_{1-x}$ films increases gradually with the Mo stoichiometry *x*, and the highest $T_c$ = 7.9 K was found in $Mo_{0.83}Si_{0.17}$, where homogeneous films with the maximum Mo content can be obtained. The thick films of $Mo_{0.83}Si_{0.17}$ show surprising degradation in which the onset of zero-resistivity is suppressed below 2 K. The thin $Mo_{0.83}Si_{0.17}$films, however, reveal robust superconductivity even with thickness $d \approx 2$ nm**.** We also characterized wide microwires based on the 2 nm thin $Mo_{0.8}Si_{0.2}$ films with wire widths 40 and 60 µm, which show single-photon sensitivity at 780 nm and 1550 nm wavelength.

## 1. Introduction

Due to the robust superconductivity, extraordinary phase homogeneity and uniformity, and little requirements on the crystal structure of the seed layer, amorphous superconducting thin films show great potential for the fabrications of superconducting micro and nano devices [1-5]. For instance, the amorphous WSi based superconducting nanowire single photon detectors (SNSPDs) have shown excellent detection performance at telecom wavelengths [1, 2]. These devices have further extended the detection capability to middle infrared photons with saturated internal quantum efficiency [6-8]. Moreover, amorphous superconductors based SNSPDs can be flexibly integrated onto the nanophotonic waveguides in traveling-wave geometry since the waveguide materials can be easily matched with these amorphous superconductors. The realization of this architecture provides a direct electronic readout of quantum information in integrated quantum photonic circuits [9].

The operation temperature of WSi SNSPDs, however, is generally limited by low temperatures (<1 K) due to the relatively low critical temperature of WSi thin films (<5 K) [1,5,10,11]. The amorphous MoGe and MoSi, which have demonstrated comparable detection performance with WSi, are reported to have critical temperatures of 7.4 and 7.5 K for

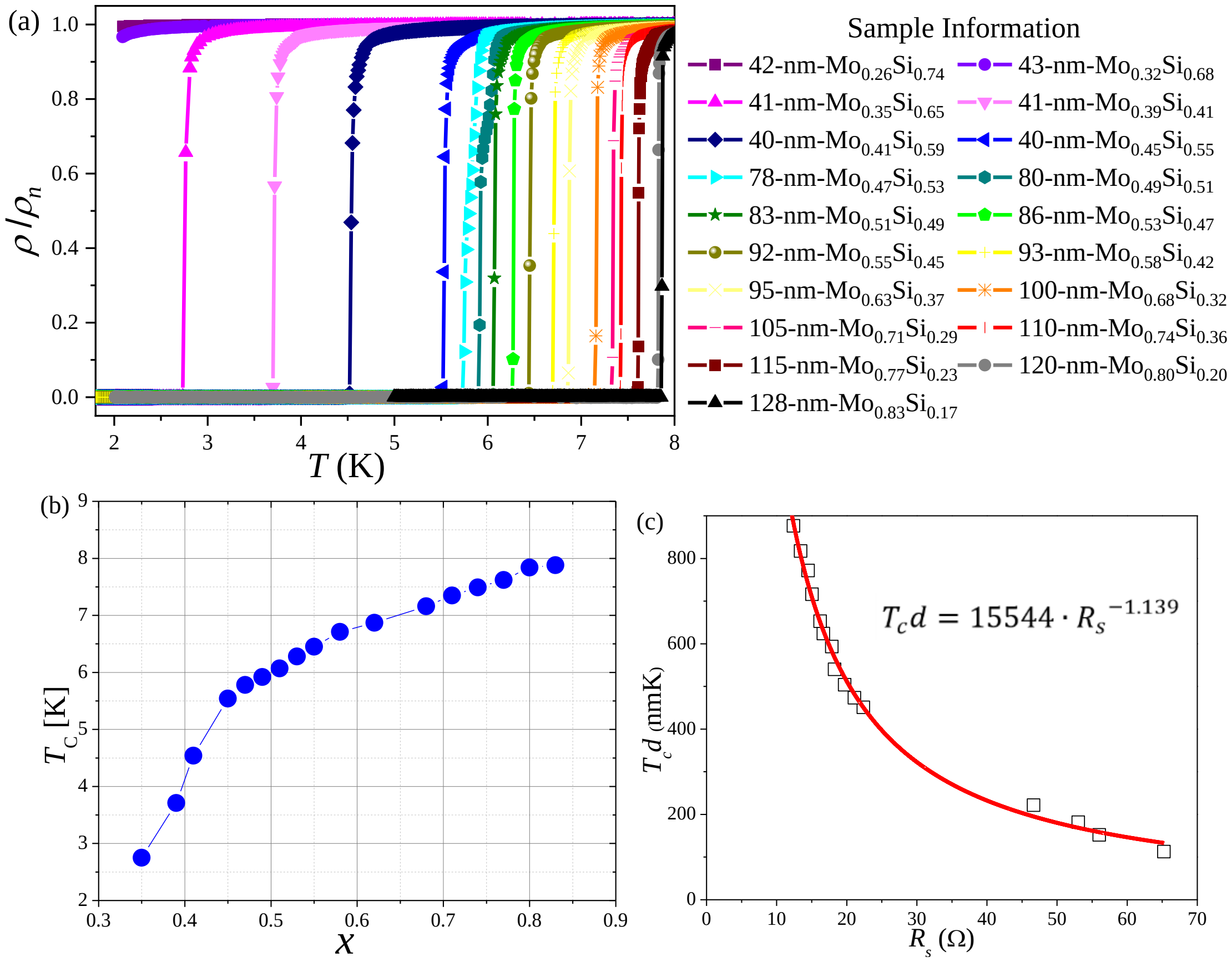

Fig.1 (a) The normalized resistivity versus temperature for thick $Mo_x\ Si_{1-x}$ films with $x$ ranging from 0.26 to 0.83. (b) The critical temperature as a function of the stoichiometry $x$. (c) $T_c\,d$ versus sheet resistance $R_S$ curve with a fit to the universal scaling law (see text) [22]. The $R_s$ was measured for 4 × 4 mm$^2$ $Mo_x\ Si_{1-x}$ films on a Si substrate with a four-wire Van der Pauw configuration.

the corresponding thin films [4,11-13]. The MoGe- and MoSi-SNSPDs are therefore able to be operated with saturated internal quantum efficiency above 2 K, which can be easily accessible with a closed-cycle cryocooler, and it has been recently reported that the system detection efficiency can be as high as to 98% [14]. Moreover, the MoSi-SNSPDs (based on 3-nm-thick MoSi film) have shown a saturated internal quantum efficiency for 1550 nm photons with a wire width up to 3 µm, where the active device area can be up to 400 × 400 µm$^2$ [15]. The high absolute bias current in MoSi wide wires reduces constraints on readout and biasing. The exploring MoSi wide detectors with width of tens micrometer is attactive for support the theory of detection in wide wires and the realization of recent proposed concept of using SNSPDs for dark matter (DM) detection where the large active area of detector (1000 µm$^2$ and larger) is requreid. Furthermore, the amorphous MoSi nanowires integrated on silicon nitride waveguides have shown saturated on-chip detection efficiency for telecom wavelength photons at a temperature of 2.1 K [9]. These fascinating results suggest that amorphous MoSi may be the preferable material for the future detector fabrication, either for fiber coupled detectors or for waveguide integrated detectors operating at higher temperature [3, 14-20].

Beyond the applications of amorphous MoSi for SNSPDs, it has been recently reported that MoSi can be successfully deposited on a flexible adhesive tape and fabricated into quantum interference devices [21]. These devices can be operated under varying flexure conditions and demonstrate robust properties, showing a great potential in quantum information technology or magnetic-field shielding with complicated geometries.

In the literature on MoSi, the reports about the superconducting properties of amorphous MoSi thin films differ considerably, and the critical temperature $T_c$ of films does not exceed the value of 7.5 K [11-13]. It has also been reported that the highest $T_c$ of MoSi films was highly dependent on the underlying substrate and the deposition temperature [12]. To systematically investigate the physical properties of amorphous MoSi films and their applications for SNSPDs fabrication, we deposited a series of MoSi thin films at ambient temperature, with thicknesses down to 1 nm, where the critical temperature of $Mo_XSi_{1-X}$ films can reach up to 7.9 K. We have performed comprehensive magnetotransport and magnetic measurements on the films

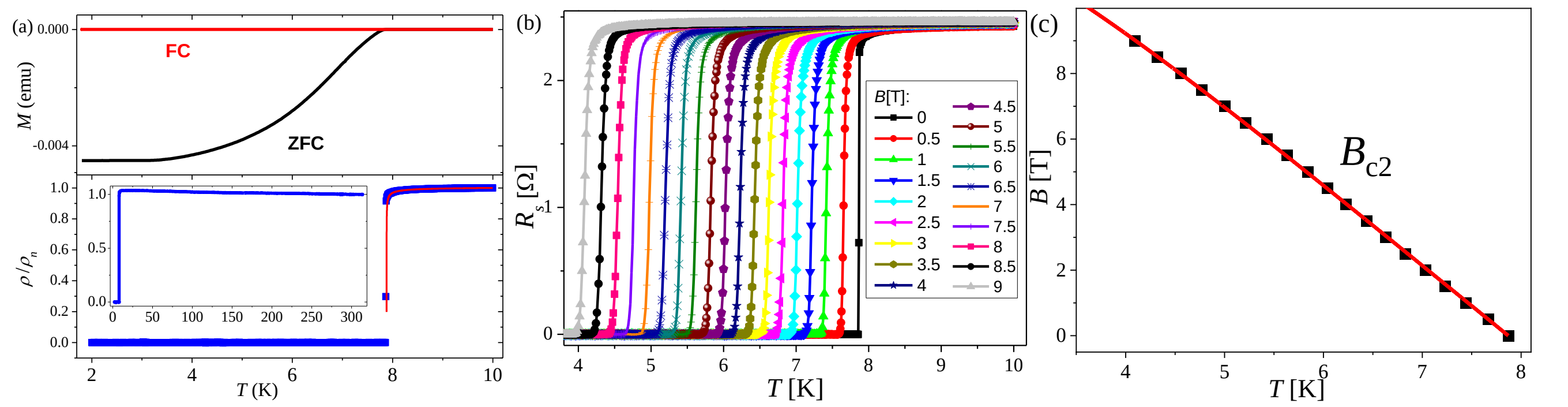

Fig.2 (a) The normal-to-superconducting transitions for a *480*-nm-thick $Mo_{0.83}Si_{0.17}$ film illustrated by magnetic moment and transport measurements. The transport measurement was performed on a 4 × 4 $mm^2$ film by using a four-wire Van der Pauw configuration, while the magnetic moment has been measured on a 3 × 4 $mm^2$ film. Inset: normalized resistivity versus temperature from 300 K to 2 K. (b) The magneto-transport measurements of the same 4 × 4 $mm^2$ film in perpendicular magnetic fields from 0 to 9 T. (c) The temperature dependence of the upper-critical field extracted from (b). The red solid line is a fit to the WHH theory.

with the most optimal stoichiometry that we used for further device fabrication. The microstructures were characterized in terms of their superconducting properties and single-photon sensitivity to photons with energy of 0.8 eV (1550 nm) at 0.3 K.

## 2. Thin films results and discussions

### *1.1 Experiments*

The $Mo_XSi_{1-X}$ films were deposited by co-sputtering from Mo and Si targets at 3 mTorr of argon pressure on silicon nitride substrates, and were then capped *in situ* with a 2 nm of amorphous Si layer. The thickness of the resulted films was inferred from the predetermined growth rate and the deposition time. To obtain highly homogenous films, we have limited the deposition rate from the Si target at relatively low power, where the preferred stoichiometry (confirmed by the energy-dispersive X-ray (EDX) spectroscopy) is adjusted by the power of Mo targets. To further guarantee the homogeneity of the resulting films, the plasma was stabilized for at least five minutes prior to each sample deposition.

The sheet resistances of the thin films were measured by using a four-wire Van der Pauw configuration. For the microbridges, the resistance has been measured by using a standard four-wire connection, and the corresponding resistivities and sheet resistances were calculated by the measured resistance and the geometries of the bridges. The transport measurements were carried out in a physical property measurement system (PPMS *Quantum Design Inc.*) under various perpendicular magnetic fields ranging from 0 up to 9 T. The magnetic properties were studied by using a magnetic properties measurement system (MPMS 3, *Quantum Design Inc.*, equipped with a reciprocating sample option (RSO)), where the fields were also perpendicular to the films.

### *1.2 Composition dependence*

We have deposited a series of $Mo_XSi_{1-X}$ films, with *x* ranging from 0.26 to 0.87. The normal state of the films shows a weakly insulating behavior. Figure 1(a) shows the

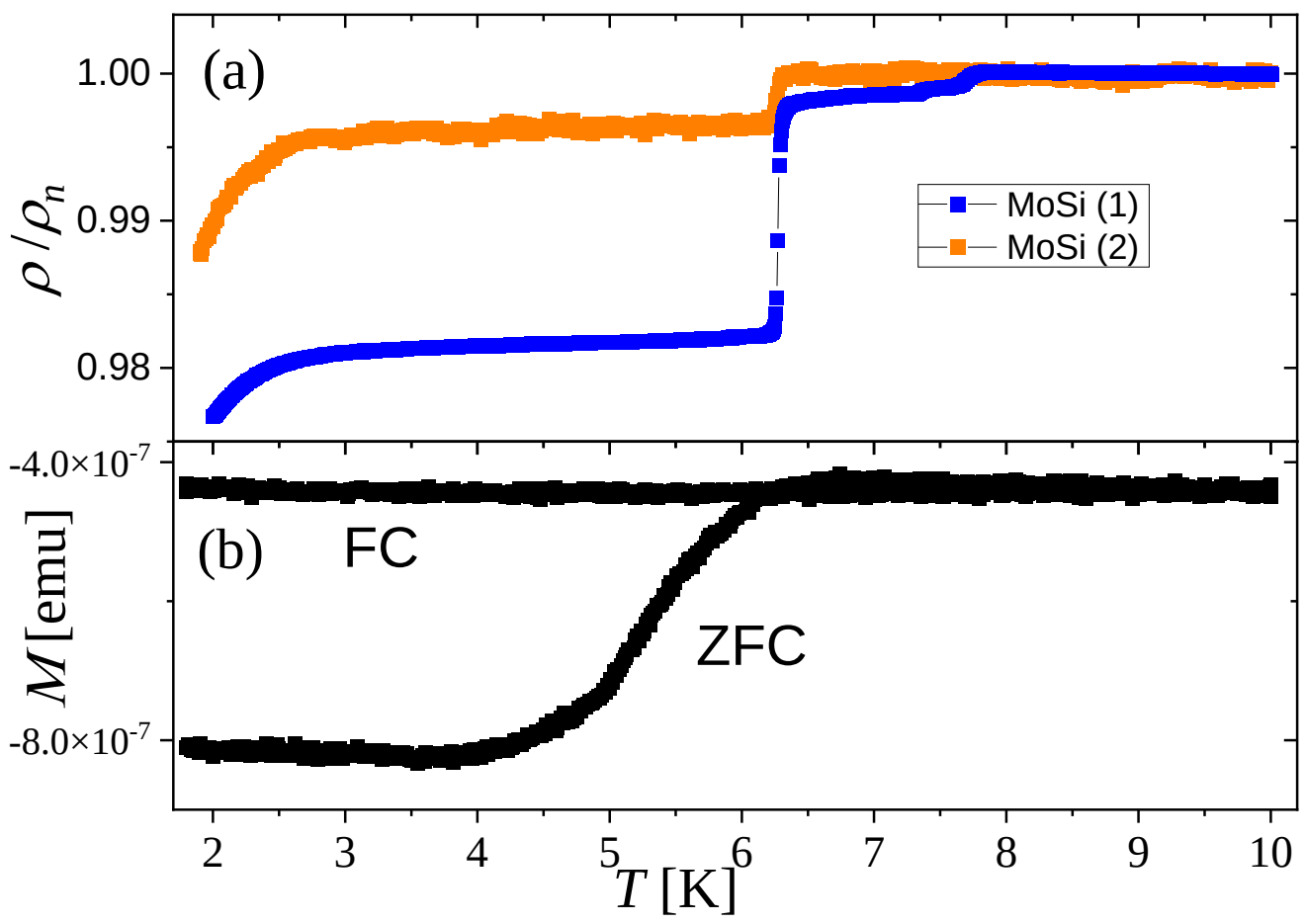

Fig.3 (a) The normalized resistivity as a function of temperature for the thick $Mo_{0.87}Si_{13}$ films. Both sample 1 and sample 2 are from the same deposition running. The resistance of sample 1 was measured on a 4 × 4 $mm^2$ film with a four-wire Van der Pauw configuration, and that of sample 2 on a 3 × 10 $mm^2$ film with a normal four-wire configuration, which was cut from a 10 × 10 $mm^2$ film. (b) The magnetic moment as a function of temperature measured for a 3 × 4 $mm^2$ film, cutting from the same film as Sample 2.

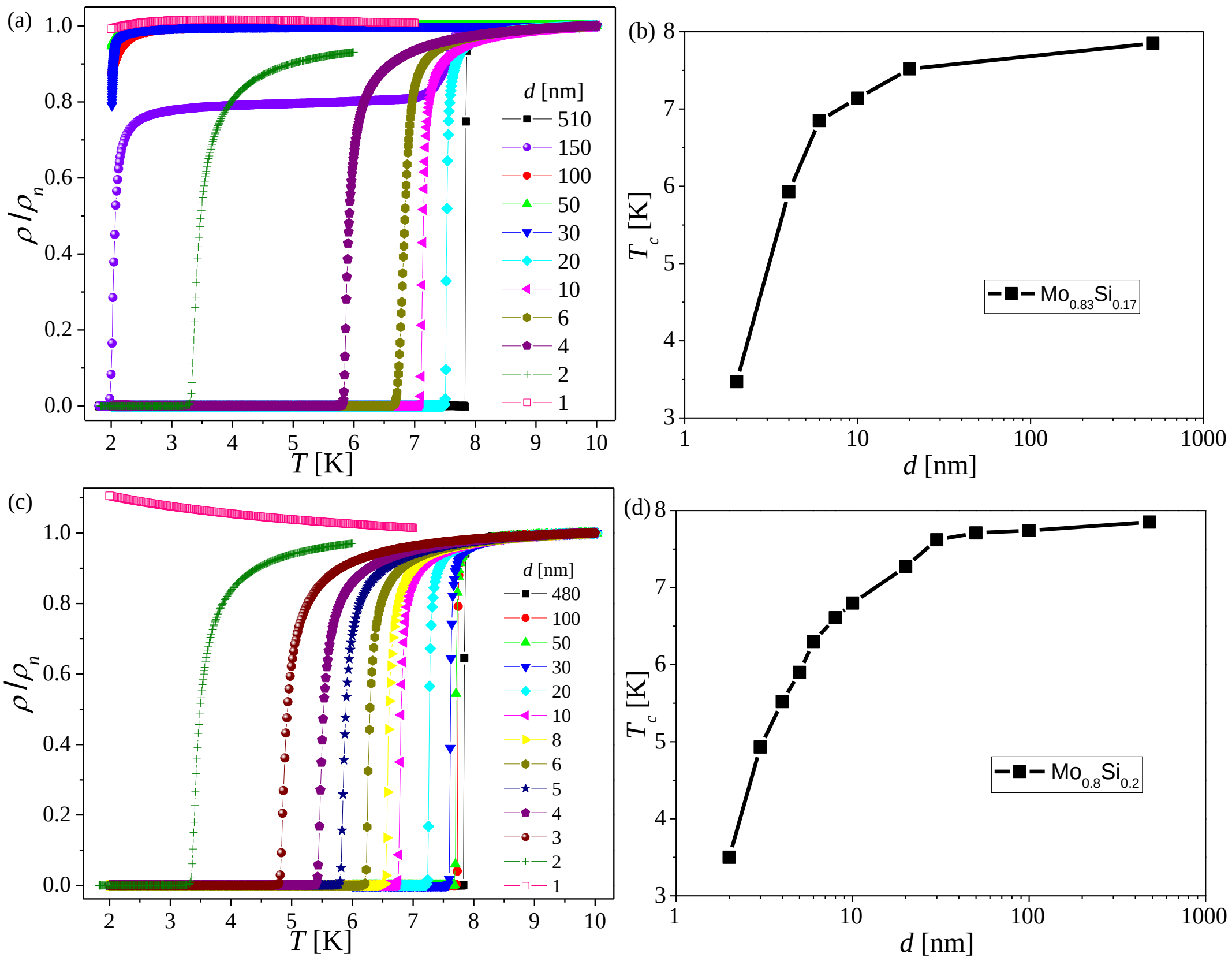

Fig. 4 (a) and (c) The normalized resistivities of $Mo_{0.83}Si_{0.17}$ and $Mo_{0.8}Si_{0.2}$ films with various thicknesses as functions of temperature. The resistivities were calculated by the measured resistance and the geometries of the 100-μm-wide bridges (with length 1000μm). The resistance was measured by using standard four-wire connections on the microbridges. (b) and (d) The critical temperatures as a functions of film thickness for $Mo_{0.83}Si_{0.17}$ and $Mo_{0.8}Si_{0.2}$, respectively. Due to the degradation in thicker $Mo_{0.83}Si_{0.17}$ films, the corresponding $T_c$(d) data are not shown.

zero-field normal to superconducting transitions for all films, with $x$ ranging from 0.26 to 0.83, where a $T_c$ above 2 K firstly appears in the 41-nm-thick $Mo_{0.35}Si_{0.65}$. The $T_c$ as a function of the specified stoichiometry is shown in Fig. 1(b). For $x < 0.45$, the transition temperatures depend strongly on the composition of the films, and gradually increase with $x$ in the range from 0.45 to 0.83.

To more generally describe the $T_c$ dependence on the film characteristics, we applied the empirical universal scaling law from Ivry *et al*.[22], in which the product of film thickness $d$ and the critical temperature $T_c$ scales as a power law of the sheet resistance of $R_S$, $T_c \times d = A \times R_S^{-B}$, where $A$ and $B$ are fitting parameters [22]. In Fig. 1(c), we summarized these properties of all films, and the solid line is the universal scaling law fit. The composition driven superconducting-to-weakly insulating transition is well described by such a universal scaling law. It is important to note that the resulting scaling exponent $B$=1.139±0.029 is fully consistent with the previous investigations on MoSi films [13], representing an intrinsic property of MoSi. Our results also suggest that the empirical scaling law could be generally applied to reveal the physical nature of superconductor-to-weakly insulator or disorder metal transitions.

The highest $T_c$ of 7.9 K was found for an as-grown 128-nm-thick $Mo_{0.83}Si_{0.17}$ film. To characterize its superconducting transition, zero-field cooling (ZFC) and field-cooling (FC) RSO magnetization, and the resistivity were measured in the temperature interval 1.8-10 K, as it is shown in Fig. 2(a). Figure 2(b) depicts the magneto-transport measurements in magnetic fields from 0 to 9 T, where $T_c$ gradually decreases from 7.9 K to 4.1 K. With increasing field, the transitions broaden slightly, and the field dependence of $T_c$ (where the resistivity drops to half of the normal state resistivity) is plotted in Fig. 2(c) as $B_{c2}(T)$. It can be well fitted by the Werthamer–Helfand–Hohenberg (WHH) model [23], resulting in a zero-temperature upper-critical field $B_{c2}(0) \approx 13.42$ T.

With higher Mo fraction, we observed the onset of superconductivity on the ρ($T$) curves, where the resistivity shows a falling down at a certain temperature. The resistivity, however, does not disappear completely, and the transitions among different samples show clear difference (Fig. 3(a)). To characterize the superconductivity in samples with $x > 0.85$, we also performed the magnetization experiment on $Mo_{0.87}Si_{0.13}$ (Fig. 3(b)). Consistent with the ρ($T$), we also observe possible diamagnetism in these samples. The absolute values of magnetic moment under the zero-field cooling, however, is found to be several orders of magnitude smaller than that in $Mo_{0.83}Si_{0.17}$ film. These results suggest that there exist superconducting domains with preformed Cooper pairs in these samples. These rare superconducting domains, however, are separated by the non-superconducting region, leading to the absence of global coherence. We therefore conclude that in films with compositions $x > 0.85$, there are signs of possible phase separations and the resulting superconductivity in such films are not homogenous amy more, as manifested in seemingly two different critical temperatures (Fig. 3). As a result, we restricticed our further investigations in films with $x < 0.85$.

## *1.3 Thickness dependence*

Generally, the application of superconducting films for device fabrications involves films with thicknesses ranging from a few nanometers to a few tens of nanometers. For instance, the thickness of SNSPDs for visible and infrared photons are commonly designed to be from 3 nm to 10 nm, but it can reach up to a few hundred nanometers for the detection of single soft X-ray photons [5,24]. We investigated here the thickness dependence of the superconducting properties of $Mo_{0.83}Si_{0.17}$ and $Mo_{0.8}Si_{0.2}$ for such practical applications.

Figure 4(a) depicts the zero-field normal-to-superconducting transitions for the $Mo_{0.83}Si_{0.17}$ films, with $d$ ranging from 2 nm to 510 nm. The thick films ($d \geq 30$ nm), however, show a surprisingly fast decrease of the transition temperature which is significantly suppressed below 2 K. Although the onset of the drop in resistivity for the 150-nm-thick film occurs already at 7.8 K, the zero resistivity is only reached around 2 K. While the thick $Mo_{0.83}Si_{0.17}$ films degrade quickly with time, the superconductivity of the thin films is found to be extremely robust. For instance, by placing the 2 nm $Mo_{0.83}Si_{0.17}$ films in air for three months, the $T_c$ only shows a slight degradation by 0.2 K. Even the 1 nm thin $Mo_{0.83}Si_{0.17}$ film still shows an onset to superconductivity at ~3.7 K. The $T_c$ for these $Mo_{0.83}Si_{0.17}$ thin films are summarized in Fig. 4(b). Because of high $T_c$ of a few nanometer thick amorphous $Mo_{0.83}Si_{0.17}$ films ($T_c$ = 6.85 and 5.93 K for 6- and 4-nm thick films), such films would still be preferable candidates for SNSPDs fabrications.

We also investigated the physical properties of $Mo_{0.8}Si_{0.2}$. These films show robust superconductivity down to 2 nm (Fig. 4(c)). The corresponding $T_c$ are plotted in Fig. 4(d), where a significant suppression on $T_c$ from the is observed for $d < 30$ nm. Although the $T_c$ for $Mo_{0.8}Si_{0.2}$ films are slightly below the $Mo_{0.83}Si_{0.17}$ peers, the more robust superconductivity makes them very suitable for device fabrications.

Based on results of the thickness dependence on $T_c$, we expect that the critical thickness for $Mo_{0.83}Si_{0.17}$ and $Mo_{0.8}Si_{0.2}$ can even be smaller than 1 nm. To describe the thickness-tuned superconducting-to-weakly insulating transitions in these films, we again applied the universal scaling law on the corresponding $T_c(d)$ data. Figure 5 shows the respective $T_c \times d(R_s)$ dependence and the fitting result from this scaling law. When expressed in units $T_c \times d(R_S)$, the $T_c(d)$ dependences of $Mo_{0.83}Si_{0.17}$ and $Mo_{0.8}Si_{0.2}$ almost coincide. The combination of the composition-driven and thickness-driven $T_c(d)$ data is shown in Fig. 5 (b) on a log-log scale. A global fit to all the data yields $A \approx$XX and $B \approx$ YY, with slight deviation of the common trend for the respective largest sheet resistances. The good scaling over two decades may hint to a truly universal behaviour of

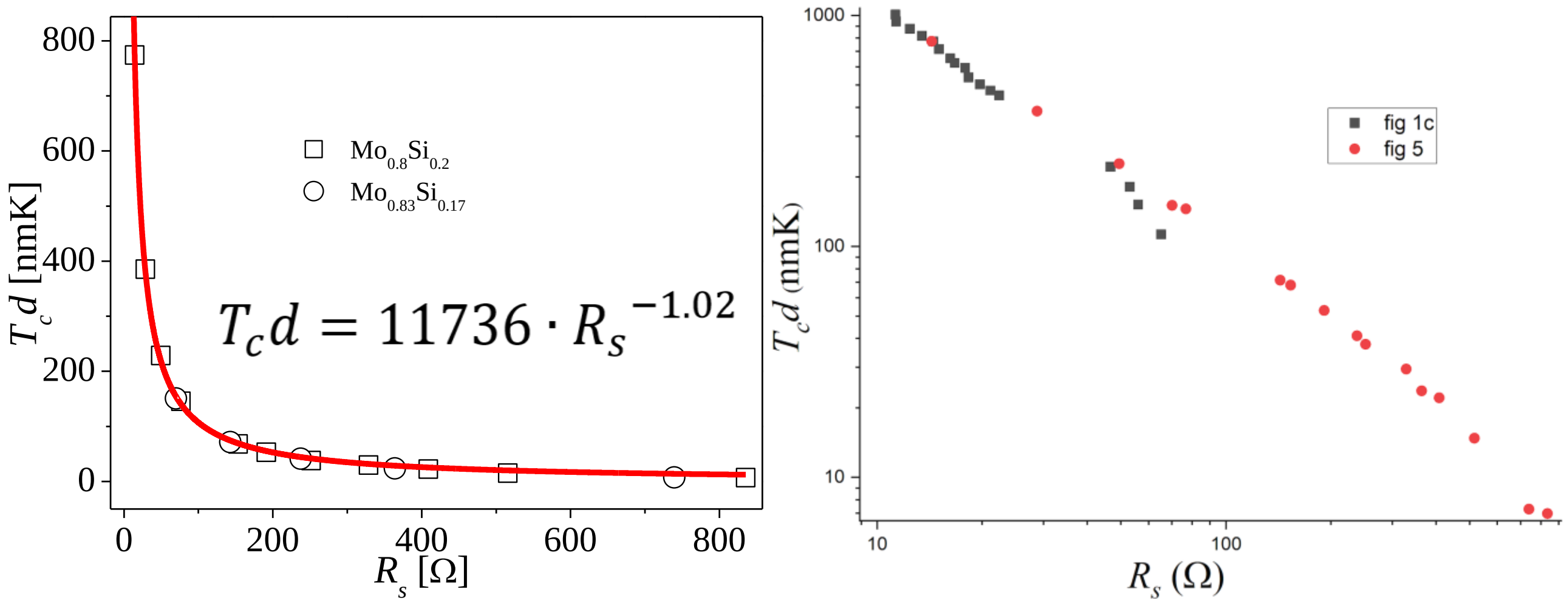

Fig. 5 (a) $T_c d$ versus $R_s$ curve with a fit to the universal scaling law for various thicknesses of $Mo_{0.83}Si_{0.17}$ and $Mo_{0.8}Si_{0.2}$ films. (b) The combination of the composition-driven and thickness-driven $T_c(d)$ data on a log-log scale, with the result of a global fit to the universal scaling law.

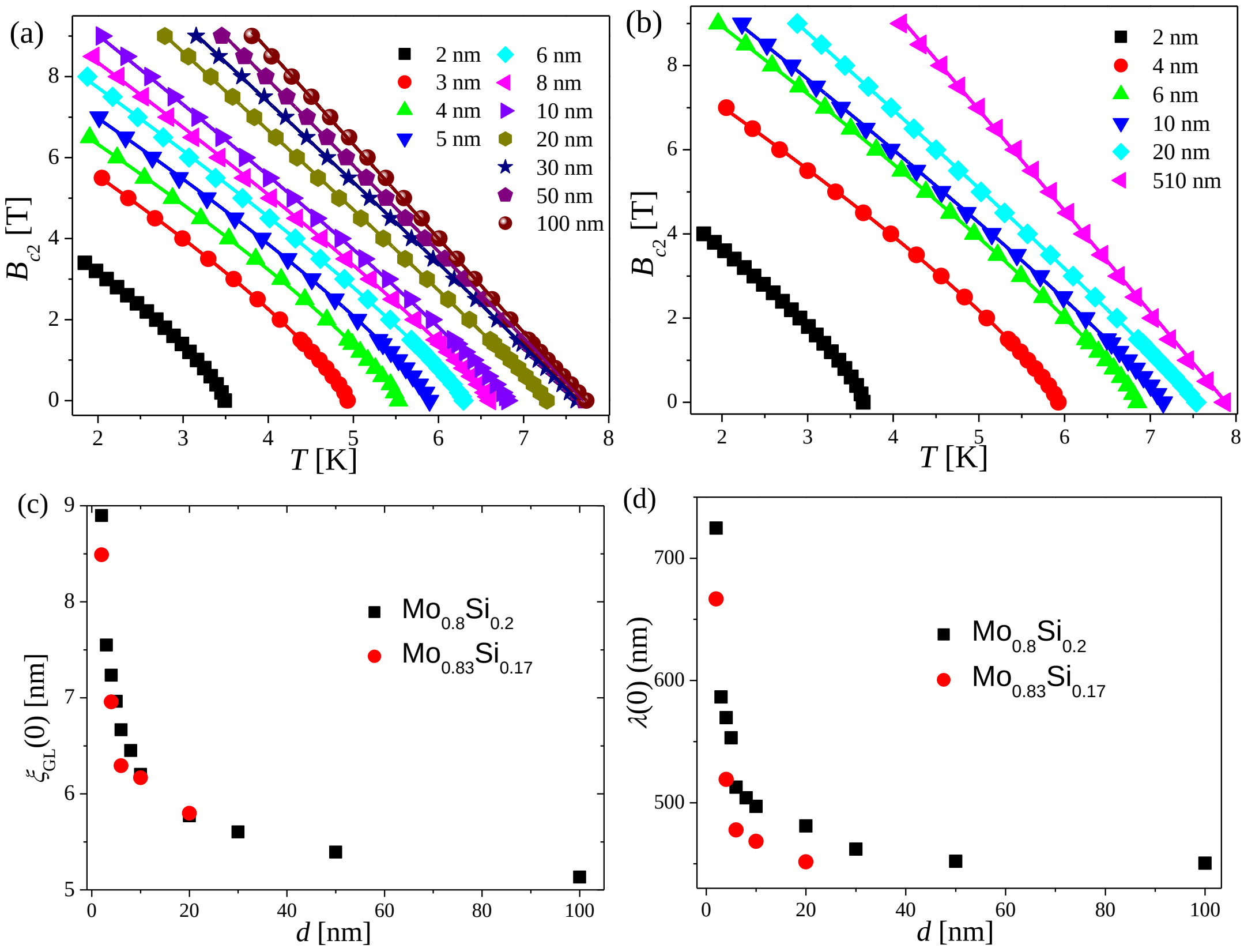

Fig. 6 (a) and (b) The upper-critical fields as functions of film thickness and temperature for $Mo_{0.8}Si_{0.2}$ and $Mo_{0.83}Si_{0.17}$, respectively. The solid lines are fits to the proposed empirical equation. (c) and (d) The zero-temperature GL coherence lengths and magnetic penetration depths as functions of film thickness for the $Mo_{0.83}Si_{0.17}$ and $Mo_{0.8}Si_{0.2}$ films.

$T_C \times d(R_S)$, irrespective of the mechanism that is responsible for the $T_c$ reduction.

To further characterize the physical properties of the thin films, we have performed magneto-transport measurements in magnetic fields up to 9 T. Figures 6 (a) and (b) summarize the temperature dependence of the upper critical field for the $Mo_{0.8}Si_{0.2}$ and $Mo_{0.83}Si_{0.17}$ films, respectively. For the bulk films, the $B_{c2}(T)$ nearly linearly depends on temperature. With decreasing film thickness, a downturn of $B_{c2}(T)$ appears near $T_c(0)$, which has been also observed in other disordered or amorphous superconducting thin films [5,25]. Despite the $B_{c2}(T)$ for the bulk films can be well fitted by the WHH theory, it is therefore not applicable to the ultra-thin films, because due to this downturn of $B_{c2}(T)$, it is impossible to obtain a reliable linear extrapolation of $B_{c2}(0)$.

To universally describe the $B_{c2}(T)$ dependence, we found that the $B_{c2}(T)$ for superconducting thin films can be well fitted by an empirical universal equation, $B_{c2}(T) = (B_{c2}(0)/0.693)\times(T_c(0) - T)^{-B}$ (red solid lines in Figs. 6 (a) and (b)). Similar to the WHH model, we here also add a pre-factor of 1/0.693 in the equation. In the bulk-film limit, the fitted result is consistent with the WHH theory and the resulting fitting parameter $B$ is close to 1. In the thinnest investigated film $d$ = 2 nm, $B$ are found to be 0.751 and 0.747 for the $Mo_{0.8}Si_{0.2}$ and $Mo_{0.83}Si_{0.17}$ films, resepctively.

From Ginzburg-Landau (GL) theory, the zero-temperature GL coherence length $\xi_{GL}(0)$ is related to the magnetic-flux quantum and the zero-temperature upper critical field, $\xi_{GL}(0) = [\Phi_0 / 2\pi \times B_{c2}(0)]^{0.5}$[5]. From the extrapolated $B_{c2}(0)$, we obtain the composition dependence of $\xi_{GL}(0)$, which is depicted in Fig. 6(c). In the dirty limit, the zero-temperature magnetic penetration depth $\lambda(0)$ is expressed as $\lambda(0) = [\hbar\rho_n / \pi\mu_0\Delta(0)]^{0.5}$, where $\hbar$ is the Planck constant, $\mu_0$ is the vacuum permeability, and $\Delta(0)$ is the zero-temperature superconducting energy gap (which can be estimated by $\Delta(0) = 1.764 k_B T_c$ [26]. The resulting thickness dependence of $\lambda(0)$ is shown in Fig. 6 (d).

## *1.4 Wire width dependence*

To characterize the homogeneity of our amorphous films, we have investigated a series of superconducting microwires (with wire widths ranging from 2 to 1000 μm) based on the 3- and 5 nm thick $Mo_{0.8}Si_{0.2}$ films. Figure 7 shows the $R_S$ as functions of wirewidth and temperature in magnetic fields from 0 to 9 T. At zero field, the $R_S(T)$ curves coincide for all wires over a 10 mm$^2$ film, demonstrating the extremely good homogeneity of the deposited films.

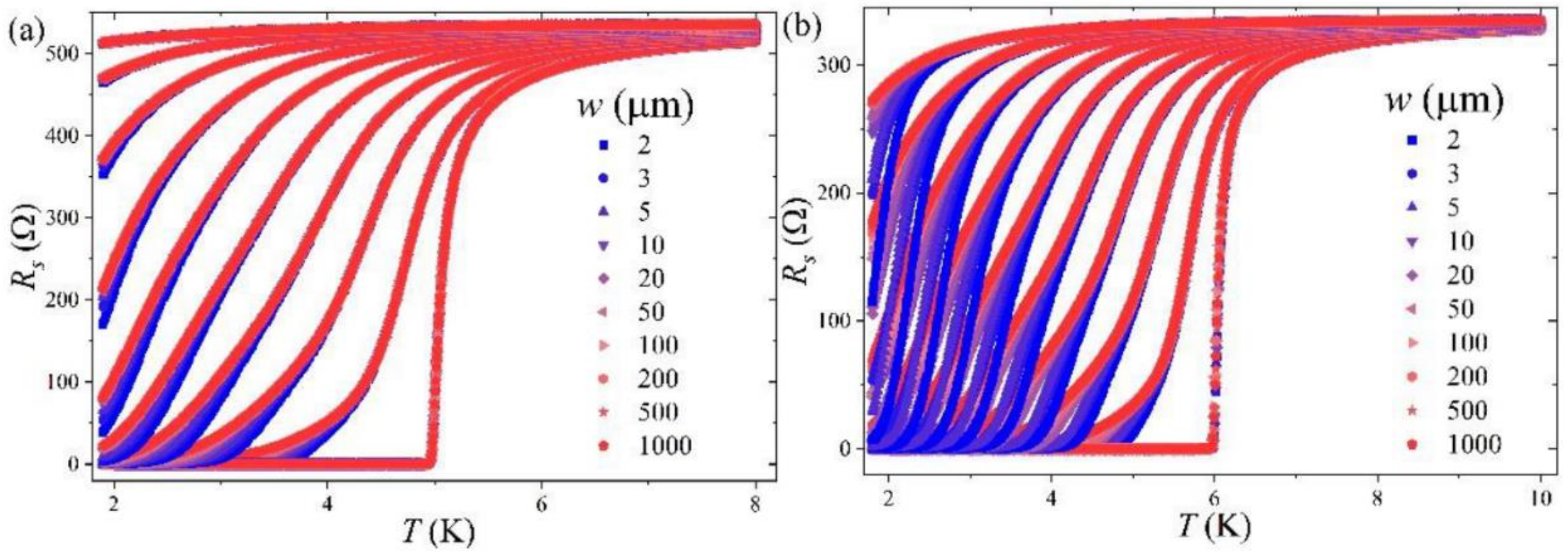

Fig. 7 The sheet resistances $R_S(B,T)$ as functions of temperature and wire width for magnetic fields ranging from 0 to 9 T in 1 T steps for 3-nm-thick (a), and 5-nm-thick $Mo_{0.8}Si_{0.2}$ (b).

With increasing magnetic field, the $R_S(B,T)$ curves are significantly broadened, and the transitions are shifted towards lower temperatures. By lowering the temperature, the $R_S(B,T)$ curves for all the wires separate in such a way that the resistivity in the narrow wires is significantly suppressed, thereby leading to a narrowing of the transition to the zero-resistance state. From our recent experiments on the width dependent magnetic-field induced superconductor-to-insulator quantum phase transitions, we conclude that this effect is due to the change of the vortex interactions in superconducting microsystems with reduced size [27,28], thereby leading to an enhanced vortex pinning that eventually suppresses the resistivity near the normal-to superconductor transition in narrow wires.

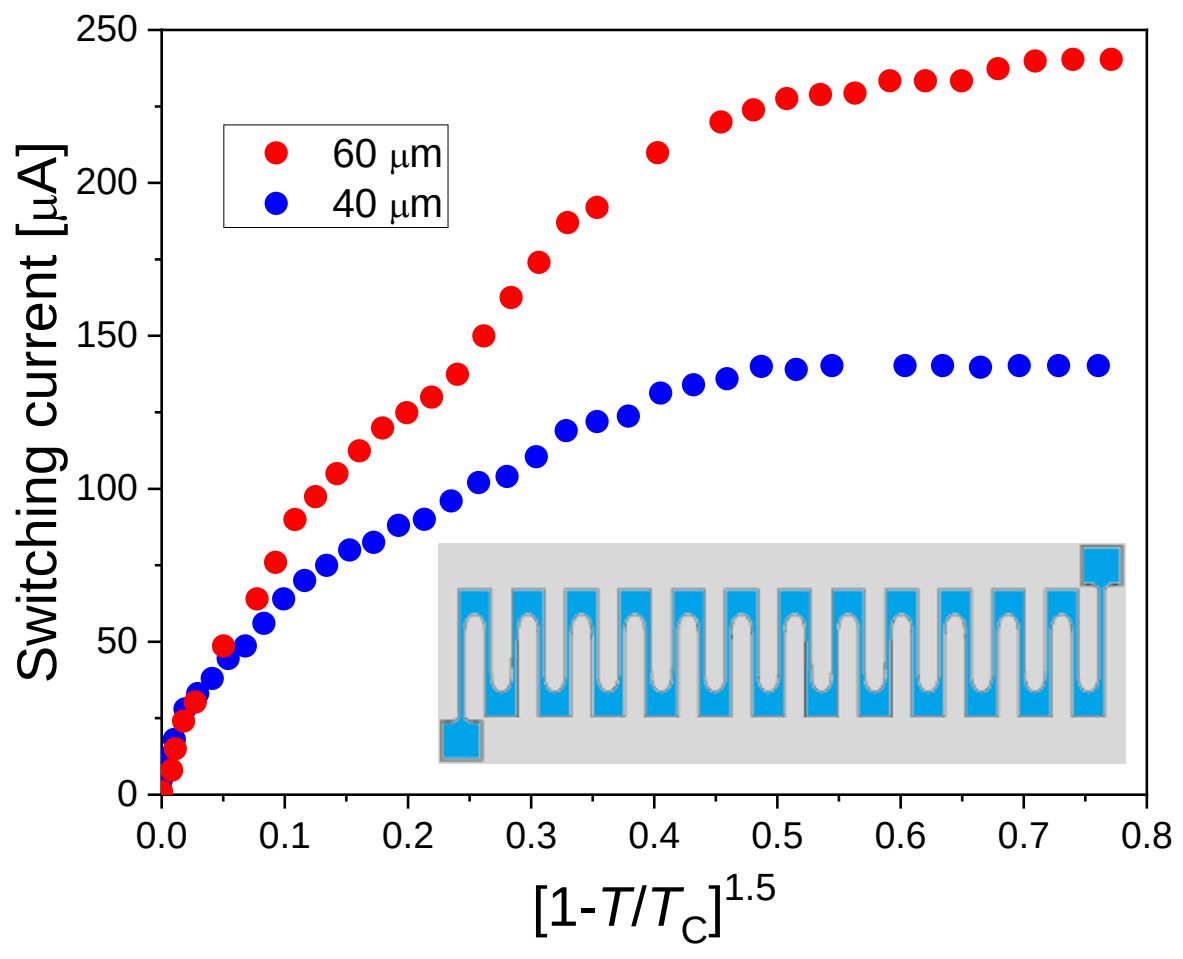

Fig. 8 The switching current as a function of temperature. The experimental data are displayed for 40 μm and 60 μm wide meanders using blue and red points respectively.

## 3. Photon detection in MoSi based microwires

The recent demonstration of single-photon detection in microscale MoSi meanders with widths of 1 and 3 μm [15] opens new doors for research of the photon detection mechanism in SNSPDs. We further explore the single-photon detection in microscale wires made of MoSi films. We chose widths of several tens microns taking the following restrictions into account: a) to reach new bounds in DM (dark matter) search using SNSPDs, detectors should be as wide as possible; b) the effect of a non-linear distribution of transport current across the wire limits the width of wire for detectors.

We herein fabricated a 40- and a 60 μm wide microwires based on 2 nm $Mo_{0.8}Si_{0.2}$ films to investigate their sensitivity to infrared photons. We should note that MoSi films for detectors were sputtered in another sputtering system than films for the material study described above. The stoichiometry of MoSi for detector fabrication was determined by X-ray diffraction (XRD) analysis (Rigaku XRD SmartLab) with attenuator correction (high resolution monochromator PB-Ge(220) × 2) on a thick 50 nm film. Despite using different deposition systems, the transport and superconducting properties of the $Mo_{0.8}Si_{0.2}$ films have been found similar for both films with the same stoichiometry and thickness. The variation of the critical temperature was found to be within 0.3 K. The sheet resistance has only 4% difference between the samples.

The design of the wide detectors is illustrated in Fig. 8. We firstly characterized the experimental switching current $I_{SW}$ for these two meander microwires in the temperature range from $T_c$ down to the operating temperature of 300 mK. The temperature dependence of the switching current is displayed in Fig.8 for 40 μm (blue points) and 60 μm (red points). Starting at 0.4 of relative temperature, the switching current of the 40-μm-wide meander is saturated, while the

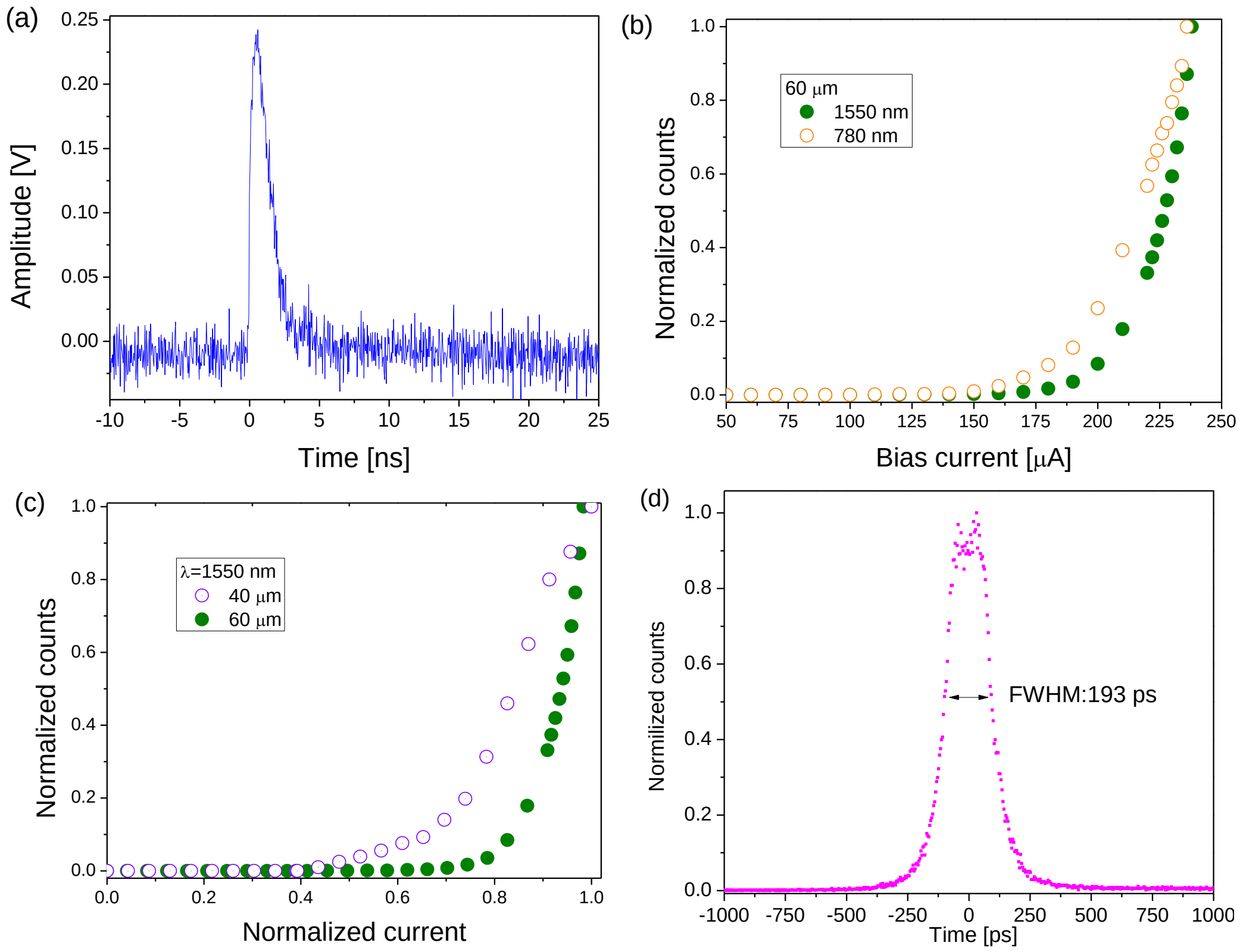

Fig. 9 (a) Example of a single voltage pulse taken with the 40 μm wide meander (b) Normalized photon counts versus absolute bias current taken on the 60 μm wide meanders upon illumination indicated in the legend; (c) Optical response of 40 μm and 60 μm wide meanders at 1550 nm wavelength. All measurements were done at 300 mK. (d) Timing jitter of 60 μm wide meander taken at 95% of switching current under illumination of 1550 nm wavelength.

60-μm-wide device shows a plateau at lower absolute temperature. The switching current was then measured to be 140 and 239 μA for 40 and 60 μm respectively. To estimate the depairing current in meanders, the temperature-dependent correction factor to the GL temperature dependence of the depairing current density in the extreme dirty limit was calculated [29]. We found the ratio between the experimental switching current and depairing current in our detectors to be 0.59 and 0.38 for 40 and 60 μm, respectively.

To optically characterize our devices, we prepared an experimental setup using a single shot type He-3 cryostat.

The detectors were first mounted on the sample holder using a contact glue. The holder was then placed on a 300-mK cold stage and was shielded to reduce the effect of background radiation on the detector noise. A room-temperature bias tee decoupled the high-frequency path from the DC bias path.

The high-frequency signal was carried out of the cryostat by stainless-steel rigid coaxial cables, while the DC bias was provided via a pair of twisted wires connected to a low-noise voltage source. The signal was amplified on the 300 K stage by a low-noise amplifier (LNA) with a total gain of 56 dB and then sent to a pulse counter. The optical single mode fiber feeds photons from the 1550-nm CW laser into the cryogenic apparatus by a vacuum feedthrough and is mounted on a stage above the detector surface. An optical attenuator was used to ensure the single-photon counting regime.

We observed voltage pulses from both 40 and 60 μm wide detectors (see example in Fig. 9a). Figure. 9(b) shows the photon count rate as a function of the absolute bias current for the 60 μm wide device taken upon illumination at 780 nm (opened orange cycles) and 1550 nm (closed green cycles). At ~150 μA (0.62$I_{SW}$), the wires start to be sensitive to photons for both energies, where the count rate grows more rapidly for the 780 nm photons. A similar result is observed in case of devices with various widths upon illumination at 1550 nm. Figure. 9(c) shows the detection performance at the telecom wavelength for both 40 μm and 60 μm wide detectors. The single-photon counting begins at relative bias currents as low as 0.4$I_{SW}$ for the 40 μm wide device, where the single-photon count rate was observed in the range of $10^5$-$10^6$ cps with extra attenuation to guarantee the single-photon detection regime. The dark count rate is limited below $10^2$ cps for all measured devices. The timing jitter was measured on 60 μm detector to be 193 ps with biasing of $95I_{SW}$ at 1550 nm wavelength.

We have also been tested wide detectors based on 2 nm thick $Mo_{0.8}Si_{0.2}$ films. Despite the absence the saturated detection efficiency, the fabricated devices showed single-photon sensitivity at the 1550-nm telecom wavelength. The choice of material stoichiometry and thickness is based on previous experiments and reports in literature. Thinner wires are expected to be more sensitive to low-energy photons. At the same time, we optimized the critical temperature and resistivity to be in an appropriate range for an experiment: a) $T_c$ should be high enough for operation at 300 mK; b) Maximize resistivity by increasing the silicon content.

A non-saturated behaviour of 40 and 60 μm wide detectors could be a consequence of the following reasons:

1) Detectors have non-optimal stoichiometry. Due to the different mechanism of the detection in nano and micro-sized wires, the best material composition for wide MoSi detectors might not be $Mo_{0.8}Si_{0.2}$. .

2) The ratio of the switching current to the depairing current is too low. The internal and geometrical defects can reduce the switching current.

3) Films are too thick. There is a certain uncertainty in the thickness measurements. A further reduction of the film thickness could enhance the sensitivity.

## Conclusions

We systematically studied the physical properties of superconducting MoSi films. The critical temperature $T_c$ increases gradually with Mo content, and the highest $T_c = 7.9$ K was found in $Mo_{0.83}Si_{0.17}$. The thick films of $Mo_{0.83}Si_{0.17}$ show a surprising degradation in terms of a suppression of $T_c$ below 2 K. The thin $Mo_{0.83}Si_{0.17}$ films, however, show robust superconductivity even with $d \lesssim 2$ nm. Microwires based on 2-nm thick $Mo_{0.8}Si_{0.2}$ films with widths of 40 μm and 60 μm show single-photon sensitivity at 780 nm and 1550 nm wavelength. This result has a particular relevance for dark-matter search where it is necessary to scale the detector area to masses large enough to probe new territory in the direct detection of sub-GeV dark matter by superconducting wires.

## Acknowledgements

The authors would like to thank J. Daley and M. Mondol of the MIT Nanostructures lab for the technical support related to electron-beam fabrication; also we would like to thank Dr. C. Settens from the Center for Materials Science and Engineering X-ray Facility for his assistance and advice on all matters related to x-ray measurements, and Phillip Keathley, Marco Turchetti, Adina Bechhofer for assistance in editing the final manuscript. Initial stages of this research were sponsored by the U.S. Army Research Office (ARO) and was accomplished under the Cooperative Agreement No. W911NF-16–2-0192. The later stages of this research were supported by the Fermi National Accelerator Laboratory, managed and operated by Fermi Research Alliance, LLC under Contract No. DE-AC02-07CH11359 with the U.S. Department of Energy, through the Office of High Energy Physics QuantISED program. This work made use of the Shared Experimental Facilities supported in part by the MRSEC Program of the National Science Foundation under award number DMR - 1419807. H.L. and D.Z. are supported by the Swiss National Foundation (grant No. 20- 175554).

## References

[1] Marsili, F., Verma, V.B., Stern, J.A., Harrington, S., Lita, A.E., Gerrits, T., Vayshenker, I., Baek, B., Shaw, M.D., Mirin, R.P. and Nam, S.W. (2013). Detecting single infrared photons with 93% system efficiency. *Nature Photonics*, 7(3), pp.210–214.

[2] Verma, V.B., Marsili, F., Harrington, S., Lita, A.E., Mirin, R.P. and Nam, S.W. (2012). A three-dimensional, polarization-insensitive superconducting nanowire avalanche photodetector. *Applied Physics Letters*, 101(25), p.251114.

[3] Verma, V.B., Korzh, B., Bussières, F., Horansky, R.D., Dyer, S.D., Lita, A.E., Vayshenker, I., Marsili, F., Shaw, M.D., Zbinden, H., Mirin, R.P. and Nam, S.W. (2015). High-efficiency superconducting nanowire single-photon detectors fabricated from MoSi thin-films. *Optics Express*, 23(26), pp.33792–33801.

[4] Verma, V.B., Lita, A.E., Vissers, M.R., Marsili, F., Pappas, D.P., Mirin, R.P. and Nam, S.W. (2014). Superconducting nanowire single photon detectors fabricated from an amorphous $Mo_{0.75}Ge_{0.25}$ thin film. *Applied Physics Letters*, 105(2), p.022602.

[5] Zhang, X., Engel, A., Wang, Q., Schilling, A., Semenov, A., Sidorova, M., Hübers, H.-W. ., Charaev, I., Ilin, K. and Siegel, M. (2016). Characteristics of superconducting tungsten silicide $W_xSi_{1-x}$ for single photon detection. *Physical Review B*, 94(17).

[6] Marsili, F., Bellei, F., Najafi, F., Dane, A.E., Dauler, E.A., Molnar, R.J. and Berggren, K.K. (2012). Efficient Single Photon Detection from 500 nm to 5 μm Wavelength. *Nano Letters*, [online] 12(9), pp.4799–4804.

[7] Chen, L., Schwarzer, D., Lau, J.A., Verma, V.B., Stevens, M.J., Marsili, F., Mirin, R.P., Nam, S.W. and Wodtke, A.M. (2018). Ultra-sensitive mid-infrared emission spectrometer with sub-ns temporal resolution. *Optics Express*, 26(12), pp.14859–14868.

[8] Verma, V.B., Lita, A.E., Korzh, B.A., Wollman, E., Shaw, M., Mirin, R.P. and Nam, S.-W. (2019). Towards single-photon spectroscopy in the mid-infrared using superconducting nanowire single-photon detectors. *Advanced Photon Counting Techniques XIII.*

[9] Häußler, M., Mikhailov, M.Yu., Wolff, M.A. and Schuck, C. (2020). Amorphous superconducting nanowire single-photon detectors integrated with nanophotonic waveguides. *APL Photonics*, 5(7), p.076106.

[10] Kondo, S. (1992). Superconducting characteristics and the thermal stability of tungsten-based amorphous thin films. *Journal of Materials Research*, 7(4), pp.853–860.

[11] Lita, A.E., Verma, V.B., Horansky, R.D., Shainline, J.M., Mirin, R.P. and Nam, S. (2015). Materials Development for

High Efficiency Superconducting Nanowire Single-Photon Detectors. *MRS Online Proceedings Library (OPL)*, 1807, pp.1–6.

[12] Bosworth, D., Sahonta, S.-L. ., Hadfield, R.H. and Barber, Z.H. (2015). Amorphous molybdenum silicon superconducting thin films. *AIP Advances*, 5(8), p.087106.

[13] Banerjee, A., Baker, L.J., Doye, A., Nord, M., Heath, R.M., Erotokritou, K., Bosworth, D., Barber, Z.H., MacLaren, I. and Hadfield, R.H. (2017). Characterisation of amorphous molybdenum silicide (MoSi) superconducting thin films and nanowires. *Superconductor Science and Technology*, 30(8), p.084010.

[14] Reddy, D.V., Reddy, D.V., Nerem, R.R., Nam, S.W., Mirin, R.P. and Verma, V.B. (2020). Superconducting nanowire single-photon detectors with 98% system detection efficiency at 1550 nm. *Optica*, 7(12), pp.1649–1653.

[15] Charaev, I., Morimoto, Y., Dane, A., Agarwal, A., Colangelo, M. and Berggren, K.K. (2020). Large-area microwire MoSi single-photon detectors at 1550 nm wavelength. *Applied Physics Letters*, 116(24), p.242603.

[16] Korneeva, Y.P., Mikhailov, M.Y., Pershin, Y.P., Manova, N.N., Divochiy, A.V., Vakhtomin, Y.B., Korneev, A.A., Smirnov, K.V., Sivakov, A.G., Devizenko, A.Y. and Goltsman, G.N. (2014). Superconducting single-photon detector made of MoSi film. *Superconductor Science and Technology*, 27(9), p.095012.

[17] Caloz, M., Korzh, B., Timoney, N., Weiss, M., Gariglio, S., Warburton, R.J., Schönenberger, C., Renema, J., Zbinden, H. and Bussières, F. (2017). Optically probing the detection mechanism in a molybdenum silicide superconducting nanowire single-photon detector. *Applied Physics Letters*, 110(8), p.083106.

[18] Li, J., Kirkwood, R.A., Baker, L.J., Bosworth, D., Erotokritou, K., Banerjee, A., Heath, R.M., Natarajan, C.M., Barber, Z.H., Sorel, M. and Hadfield, R.H. (2016). Nano-optical single-photon response mapping of waveguide integrated molybdenum silicide (MoSi) superconducting nanowires. *Optics Express*, 24(13), pp.13931–13938.

[19] Caloz, M., Perrenoud, M., Autebert, C., Korzh, B., Weiss, M., Schönenberger, C., Warburton, R.J., Zbinden, H. and Bussières, F. (2018). High-detection efficiency and low-timing jitter with amorphous superconducting nanowire single-photon detectors. *Applied Physics Letters*, 112(6), p.061103.

[20] Korneeva, Yu.P., Manova, N.N., Florya, I.N., Mikhailov, M.Yu., Dobrovolskiy, O.V., Korneev, A.A. and Vodolazov, D.Yu. (2020). Different Single-Photon Response of Wide and Narrow Superconducting $Mo_xSi_{1-x}$ Strips. *Physical Review Applied*, 13(2).

[21] Suleiman, M., Torre, E.G.D. and Ivry, Y. (2020). Flexible Amorphous Superconducting Materials and Quantum Devices with Unexpected Tunability. *arXiv:2002.10297 [cond-mat, physics:quant-ph]*. Available at: https://arxiv.org/abs/2002.10297.

[22] Ivry, Y., Kim, C.-S., Dane, A.E., De Fazio, D., McCaughan, A.N., Sunter, K.A., Zhao, Q. and Berggren, K.K. (2014). Universal scaling of the critical temperature for thin films near the superconducting-to-insulating transition. *Physical Review B*, 90(21).

[23] Helfand, E. and Werthamer, N.R. (1966). Temperature and Purity Dependence of the Superconducting Critical Field,Hc2. II. *Physical Review*, 147(1), pp.288–294.

[24] Zhang, X., Wang, Q. and Schilling, A. (2016). Superconducting single X-ray photon detector based on W0.8Si0.2. *AIP Advances*, 6(11), p.115104.

[25] Semenov, A., Günther, B., Böttger, U., Hübers, H.-W. ., Bartolf, H., Engel, A., Schilling, A., Ilin, K., Siegel, M., Schneider, R., Gerthsen, D. and Gippius, N.A. (2009). Optical and transport properties of ultrathin NbN films and nanostructures. *Physical Review B*, 80(5).

[26] Bardeen, J., Cooper, L.N. and Schrieffer, J.R. (1957). Theory of Superconductivity. *Physical Review*, 108(5), pp.1175–1204.

[27] Zhang, X., Lita, A.E., Smirnov, K., Liu, H., Zhu, D., Verma, V.B., Nam, S.W. and Schilling, A. (2020). Strong suppression of the resistivity near the superconducting transition in narrow microbridges in external magnetic fields. *Physical Review B*, 101(6).

[28] Zhang, X., Lita, A. E., Liu, H., Verma, V. B., Zhou, Q., Nam, S. W., & Schilling, A. (2021). Size dependent nature of the magnetic-field driven superconductor-to-insulator quantum-phase transitions. Communications Physics, 4(1), 1-9.

[29] Charaev, I., Silbernagel, T., Bachowsky, B., Kuzmin, A., Doerner, S., Ilin, K., Semenov, A., Roditchev, D., Vodolazov, D.Yu. and Siegel, M. (2017). Proximity effect model of ultranarrow NbN strips. *Physical Review B*, 96(18).